\documentclass[12pt,a4paper]{article}
\usepackage{jcappub}
\usepackage{bm}
\usepackage{indentfirst}
\usepackage{amsmath}
\usepackage{graphicx}
\usepackage{float}
\usepackage{amssymb}
\usepackage{subfigure}
\usepackage{hyperref}
\usepackage{array}
\usepackage{amsthm}
\usepackage{mathrsfs}
\usepackage{color}
\usepackage{comment}
\usepackage[normalem]{ulem}
\usepackage{url} 
\hypersetup{
	colorlinks=true,
	linkcolor=red,
	citecolor=blue,
}
\allowdisplaybreaks[2]

\begin{document}


\title{Improved bounds on the bosonic dark matter with pulsars in the Milky Way}

\author[a]{Dicong Liang,}
\author[a,b,1]{Lijing Shao\note{Corresponding author.}}

\affiliation[a]{Kavli Institute for Astronomy and Astrophysics, Peking
University, Beijing 100871, China}
\affiliation[b]{National Astronomical Observatories, Chinese Academy of
Sciences, Beijing 100012, China}

\emailAdd{dcliang@pku.edu.cn}
\emailAdd{lshao@pku.edu.cn}

\keywords{dark matter, neutron star, Galactic motion, cross section}

\abstract{ Neutron stars (NSs) can be used to constrain dark matter (DM) since a
NS can transform into a black hole (BH) if it captures sufficient DM particles
and exceeds the Chandrasekhar limit.  We extend earlier work and for the first
time take into account the Galactic motion of individual NSs, which changes the
amount of the captured DM by as large as one to two orders of magnitude.  We
systematically apply the analysis to 413 NSs in the Milky Way, and constrain the
DM particle mass and its interaction with nucleon simultaneously.  We find that
the most stringent bound is placed by a few NSs and the bound becomes stronger after considering the Galactic motion. 
The survival of observed NSs already
excludes a cross section  $\sigma_{nX}\gtrsim 10^{-45} \, {\rm cm}^2$ for DM
particles with mass from $100\, {\rm MeV}$ to $10^3 \, {\rm GeV}$.  Especially
for a mass around $10 \, {\rm GeV}$, the constraint on the cross section is as
stringent as $\sigma_{nX}\lesssim 10^{-49} \, {\rm cm}^2$.  }

\maketitle

\section{Introduction}

According to contemporary observations, there is a large number of invisible
matter, which constitutes about 26\% of the total energy of our Universe
\cite{Planck:2018vyg}.  This kind of matter does not interact with the
electromagnetic field, but manifests itself by the gravitational effect, thus it
is called dark matter (DM) \cite{Bertone:2004pz, Clowe:2006eq}.
Weakly-interacting massive particle (WIMP), with mass in the range from GeV to
TeV, is one of the best motivated candidates of DM \cite{Jungman:1995df,
Feng:2010gw}.  In 1980s, the possible capture of WIMPs by the Sun and the Earth
was studied. If WIMPs annihilate inside the Sun or the Earth, it would produce
jets of neutrinos that might be detected by neutrino
detectors~\cite{Press:1985ug, Gould:1987ju, Gould:1987ir, Gould:1987ww}.  Later,
the effects of DM particle annihilation and kinetic heating inside neutron stars
(NSs) were discussed~\cite{Kouvaris:2007ay, Bertone:2007ae, Kouvaris:2010vv,
McCullough:2010ai, Bramante:2017xlb, Baryakhtar:2017dbj, Raj:2017wrv,
Bell:2018pkk, Camargo:2019wou, Acevedo:2019agu,Maity:2021fxw,Nguyen:2022zwb, Chatterjee:2022dhp}.  After being captured in a NS,
DM can change the inner structure of the star \cite{Sandin:2008db,
Perez-Garcia:2010xlt, Leung:2012vea, Li:2012qf, Perez-Garcia:2013dwa,
Goldman:2013qla, Xiang:2013xwa, Rezaei:2016zje, Mukhopadhyay:2016dsg,
Ellis:2018bkr, Das:2020vng, Das:2020ptd, Kain:2021hpk, Gleason:2022eeg,Shakeri:2022dwg} and
leave imprint on gravitational waves~\cite{Ellis:2017jgp, Nelson:2018xtr,
Kopp:2018jom, Das:2018frc, Alexander:2018qzg, Quddus:2019ghy, Horowitz:2019aim,
Das:2021wku, Das:2021yny, Bezares:2019jcb, Das:2020ecp, Dengler:2021qcq,
Collier:2022cpr,Karkevandi:2021ygv}.

If the captured DM concentrates in a very dense region, it can even form a
minuscular black hole (BH) inside the NS \cite{Goldman:1989nd, Gould:1989gw}.
Then the small BH can consume the star and finally produce a BH with mass around
1 to 2 solar masses \cite{Kouvaris:2018wnh, Dasgupta:2020mqg, Garani:2021gvc,
Bhattacharya:2023stq}.  Such a scenario happens only when the NS accretes a
large amount of DM, which depends on the interaction cross section  between the
DM particles and the nucleons \cite{Guver:2012ba}.  Recently, more progress was
made in improving the model of DM capture~\cite{Lopes:2020dau, Bell:2020obw,
Bell:2020jou, Bell:2020lmm, Anzuini:2021lnv, Bose:2022ola, Dasgupta:2019juq, Dasgupta:2020dik}.  After being captured, the DM
particles lose energy and then thermalize with the NS via repeatedly scattering
with nucleons \cite{Bertoni:2013bsa, Garani:2020wge}.  For asymmetric DM which
does not self-annihilate \cite{Petraki:2013wwa,Zurek:2013wia}, it accumulates in
the core of the NS after thermalization, eventually reaches the Chandrasekhar
limit, and collapses into a BH \cite{deLavallaz:2010wp, McDermott:2011jp,
Kouvaris:2010jy}.  The nascent BH grows when its accretion outstrips its Hawking
radiation evaporation \cite{Kouvaris:2012dz, Fan:2012qy, Kouvaris:2013kra,
East:2019dxt}, and finally destroys the NS.  

If DM particles are bosons, they can form a Bose-Einstein condensate (BEC),  and
the self-gravity starts for a smaller number of particles, making the
gravitational collapse easier to happen \cite{Kouvaris:2011fi}.  Including the
relativistic effects of the NS's gravitational field on the BEC transition, the
bounds on the cross section can be further strengthened \cite{Jamison:2013yya}.
If DM particles are fermions, the gravitational collapse happens only when the gravity overcomes the Fermi degeneracy pressure.  Taking into account
Yukawa-type attractive interactions, the number of DM particles necessary for
collapse  significantly decreases \cite{Kouvaris:2011gb}. 
While it is argued in Refs.~\cite{Gresham:2018rqo,Garani:2022quc} that after considering the relativistic effects, the Chandrasekhar limit does not significantly change.
Besides, DM self-annihilation and co-annihilation with nucleons can release the bounds on DM
\cite{Bramante:2013hn, Bramante:2013nma, Bell:2013xk}.  A detailed study for
both bosonic and fermionic DM can be found in Ref.~\cite{Garani:2018kkd}.
Constraining DM via black hole formation in non-compact objects can be found in Ref.~\cite{Ray:2023auh}.

The DM capture rate by a NS depends on the local DM density.  As the DM halo
density profile is not uniform, the constraint discussed above is dependent on
the position of NSs \cite{Bramante:2014zca}.  In literature, NSs were all
assumed as static in the Milky Way. But in reality, they are moving around the
center of the Galaxy, and some of them can have a large orbital eccentricity,
with PSR~J1909$-$3744 being an example (cf.~Fig.~11 in Ref.~\cite{Liu:2020hkx}).
Therefore, some of these NSs have passed the inner Galactic region during the
motion, where the DM density is much higher than their current location.  In
this paper, we for the first time consider how the Galactic motion of the NSs
affects their accretion of DM and then affects the derived bound on the
DM-nucleon cross section.  We use the GalPot package\footnote{
\url{https://github.com/PaulMcMillan-Astro/GalPot/tree/master}}
\cite{2016ascl.soft11006M}  to calculate the orbits of the NSs, based on the
mass model of the Milky Way \cite{Dehnen:1996fa, McMillan:2011wd,
2017MNRAS.465...76M}.  We find that after taking into account the Galactic
motion, the amount of captured DM varies. For some cases, the change is as large
as one to two orders of magnitude, compared to their corresponding static cases.
In this work we use bosonic DM to illustrate, and the analysis will be applied
to fermionic DM in the future.

The paper is organized as follows.  In Sec.~\ref{DMcapture}, we calculate the
orbits of NSs in the Milky Way and compare the amount of captured DM with the
static cases.  Then, we discuss details about the mechanism for DM to form a BH
inside the NS in Sec.~\ref{BHformation}.  Finally, we use a catalog of NSs to
place improved constraint on the DM, and discuss the results in
Sec.~\ref{discussions}.

\section{Dark Matter Capture}
\label{DMcapture}

We assume that NSs have mass $M_{\rm NS} \simeq 1.5 \, M_\odot$, radius $R_{\rm
NS} \simeq 11 \, {\rm km}$, inner temperature $T_{\rm NS} \simeq 10^6  \,  {\rm
K}$, and  a uniform density $\rho_B = M_{\rm NS} / \big(\frac{4}{3}\pi
R_{\rm NS}^3\big)$.  The escape velocity at the surface of a NS is $v_{\rm
esc}=\sqrt{2 G M_{\rm NS} /R_{\rm NS}}$.  We denote $m_X$ the mass of DM
particle, and $\sigma_{nX}$ the cross section of the interaction between 
DM particle and neutron.  We consider that the velocity of DM particles
follows the Maxwell-Boltzmann distribution, and the velocity dispersion is
$\bar{v}_X$.

A DM particle is captured when it loses enough energy after scattering with
nucleon.  Taking relativistic effects into account, the capture rate is given by
\cite{McDermott:2011jp, Bramante:2013nma, Bramante:2014zca},
\begin{align}
C_{X,0} =& \sqrt{\frac{6}{\pi}} \left(\frac{\rho_{\rm DM}}{\bar{v}_X} \right)\frac{\xi N_B v^2_{\rm esc}}{m_X} \left[1-\frac{1-\exp(-B^2)}{B^2} \right]f(\sigma_{nX}) \, .
\label{capture}
\end{align}
Here $N_B$ is the total number of neutrons within the star, $m_B$ is the mass of
the neutron, $\rho_{\rm DM}$ is the density of DM.  The factor
\begin{equation}
    B^2=\frac{v_{\rm esc}^2}{\bar{v}_X^2}\frac{6 m_X m_B}{(m_X-m_B)^2} \, ,
\end{equation}
accounts for the minimum energy loss necessary to capture the DM
\cite{McDermott:2011jp}, and the parameter $\xi= \min \{ 
\sqrt{2}m_r v_{\rm esc}/p_F ,1\} \simeq \min \big\{m_X/(0.2\ {\rm
GeV}),1 \big\}$ accounts for the suppression in the capture rate due to the
Pauli blocking \cite{McDermott:2011jp,Bell:2013xk,Bramante:2014zca}. Here, $m_r$ is the dark matter-neutron reduced mass and $p_F$ is the Fermi momentum. 
The function $f(\sigma_{nX}) =
\sigma_{\rm sat} \big[1- e^{-\sigma_{nX}/\sigma_{\rm sat}}\big]$ with
$\sigma_{\rm sat}= R_{\rm NS}^2/\big(0.45 N_B \xi\big) $ gives the probability
that a DM particle scatters  \cite{Bramante:2013nma}.

In literature, NSs are treated as static, but in fact their Galactic motion cannot be neglected during their whole lifetime. When we consider the Galactic motion of NSs, $\rho_{\rm DM}$ is no longer constant for each NS but varies with time, that is, $\rho_{\rm DM}=\rho_{\rm DM}(\bm{r}(t)) $.  To reconstruct the Galatic motion of NSs, we make use of the data in the Australia
Telescope National Facility (ATNF) pulsar catalogue\footnote{
\url{https://www.atnf.csiro.au/people/pulsar/psrcat/}} \cite{Manchester:2004bp}
to set up the initial conditions, i.e.\ the position and velocity of the stars.
For most of the NSs, the proper motion in right ascension and declination, namely, the transverse velocity can be
derived from timing observations. While the radial velocity $v_r$ is difficult to measure, thus, it remains unknown for most of the NSs.  We select 413 NSs from the ATNF catalogue, whose sky location, proper motion and distance are known.  Then, we consider three configurations where we
assume $v_r=0$, $-100 \, \rm{ km/s}$, and $100 \, \rm{ km/s}$, respectively,  so that we have the full three-dimension velocity.
With the Galactic gravitational potential provided by
McMillan~\cite{2017MNRAS.465...76M}  and the position and velocity of the stars, we use GalPot package
\cite{2016ascl.soft11006M} to integrate the Galactic motion of these NSs
backward in time for the duration of their characteristic age, $t_{\rm NS}$.
The parameter $t_{\rm NS}$ in the ATNF catalog is estimated by the spin-down of
the star \cite{Manchester:2004bp}, and it might be overestimated, especially for
the millisecond pulsars that have undergone the {\it recycling} processes
\cite{Tauris:2012jp}.   
Thus, for those NSs whose characteristic age $t_{\rm NS}> 10 \, {\rm Gyr}$, we set their age to be $t_{\rm NS}= 10 \, {\rm Gyr}$.  

\begin{figure}
 \includegraphics[width=\linewidth]{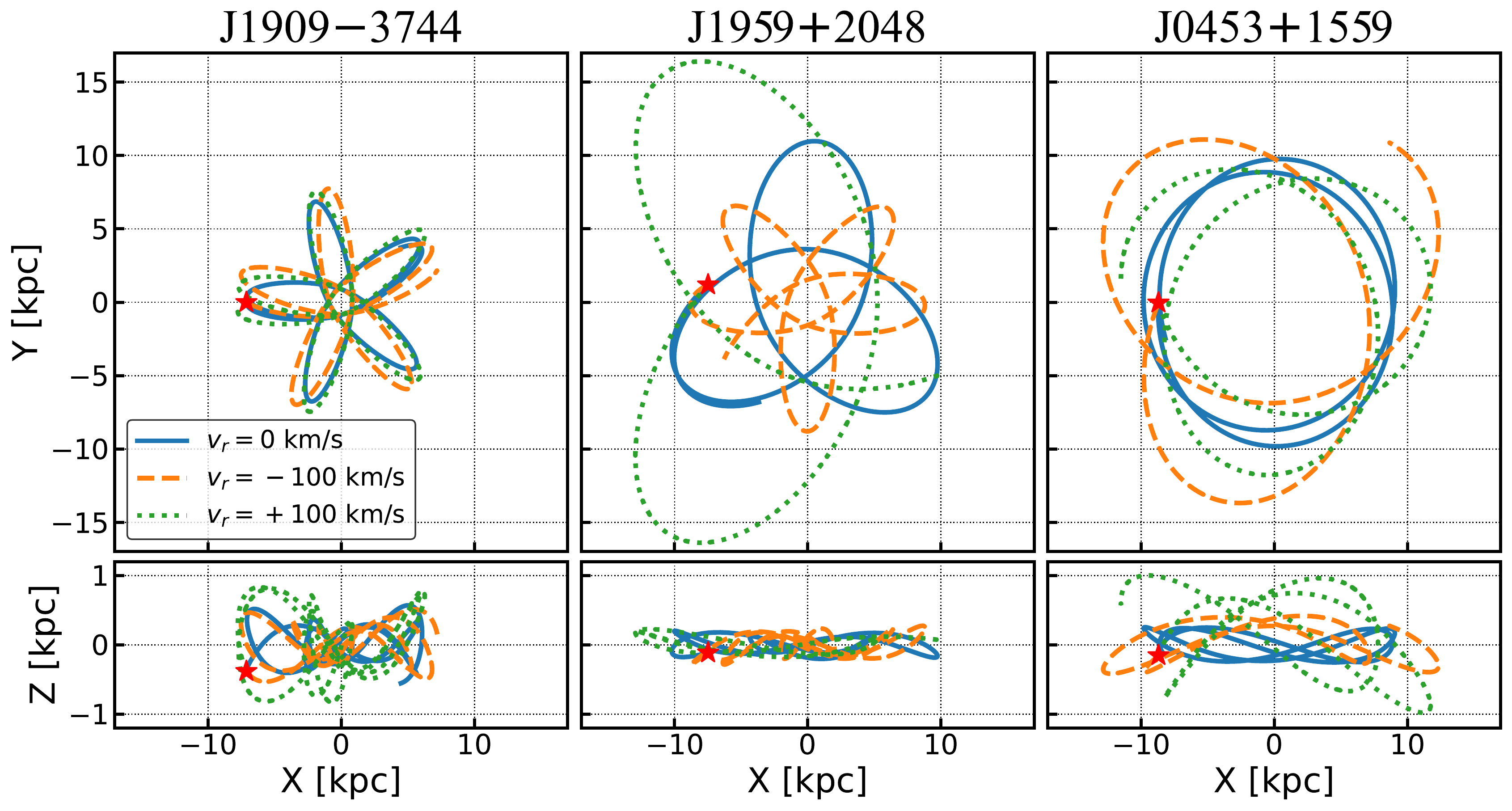}
 \caption{Galactic motion of three pulsars in the past 500\,Myr. The red star denotes the current position of the pulsars.}
\label{Galmotion}
\end{figure}

Galactic motions are calculated for all the selected 413 NSs.  Taking three NSs,
PSRs~J1909$-$3744, J1959$+$2048 and J0453$+$1559 as examples, we show their
Galactic motion in Fig.~\ref{Galmotion}.  To make the trajectory more clear, we
only show the motion in the past 500\,Myr.  Following Liu et
al.~\cite{Liu:2020hkx}, the coordinate system is chosen such that the origin is
the Galactic center and the X-Y plane coinciding with the Galactic plane.  For
PSR~J1909$-$3744, the shape of the orbit does not change much with respect to
different radial velocity, $v_r$. While for PSR~J1959+2048, the orbits are
significantly different from each other when $v_r$ is different. PSR~J0453+1559
is an example of small orbital eccentricity.

When calculating the orbit, we have made use of the best fitting Galactic potential model of the Milky Way \cite{2017MNRAS.465...76M}, in which the DM density distribution is modeled by the Navarro-Frenk-White profile \cite{Navarro:1996gj} ,
\begin{equation}
    \rho_{\rm DM} (r)=\frac{\rho_{0,h}}{ (r/r_h)^\gamma (1+3r/r_h)^{3-\gamma}  } \,,
\end{equation}
where $\gamma=1$, and  $r_h=18.6 \, {\rm kpc}$ is the scale radius. The
parameter $\rho_{0,h}$ is a normalized factor such that the local DM density,
namely the density at the Solar system, is $\rho_{\rm DM}(r=8.20 \, {\rm kpc})
=0.38 \, {\rm GeV}/{\rm cm}^3$.

In addition, when we take into account the relative motion between the NS and the DM halo, there is correction factor $\zeta$ on the capture rate, as was indicated in Refs.~\cite{Gould:1987ir,Bell:2020jou}:
\begin{align}
	C_X = \zeta C_{X,0} \, ,
\label{NewCapture}
\end{align}
where 
\begin{align}
  \zeta=& \left[1-\frac{1-\exp(-B^2)}{B^2} \right]^{-1} \bigg\{
  \frac{(B_+ B_- -1/2)[\chi(-\eta,\eta) -\chi(B_-,B_+) ] 
  }{2\eta B^2} \nonumber \\
  & + \frac{B_+ \exp(-B_-^2)/2 -B_-\exp(-B_+^2)/2 
  -\eta \exp(-\eta^2)}{2\eta B^2}  \bigg\} ,
\end{align}
$\chi(a,b)\equiv \int_a^b \exp(-y^2)dy =\sqrt{\pi}[{\rm erf}(b)-{\rm erf}(a)]/2$, $B_\pm \equiv B \pm \eta$ and $\eta=\sqrt{3/2}v_{\star}/\bar{v}_X $.
Here, $v_{\star}$ is the relative velocity between the NS and the DM halo.
When $\eta\to 0$, i.e. the NS is static relative to the DM halo, then $\zeta \to 1$. In addition, we have 
\begin{align}
    \zeta \to \zeta_{\infty}\equiv \frac{\sqrt{\pi} \, {\rm erf}(\eta) }{2 \eta},
\end{align}
when $B \gg 1$.
At the same time, we also consider that the velocity dispersion of DM $\bar{v}_X$ varies with different radius from the Galactic center. 
Here we adopt the so-called standard halo model and assume the velocity dispersion is isotropic, then we have \cite{Green:2010gw,Catena:2011kv,Strigari:2012acq}
\begin{equation}
    \bar{v}_X =\sqrt{3/2} v_c ,
\end{equation}
where $v_c = v_c (r(t))$ is the circular velocity at the radius $r$ from the Galactic center. 
Notice that, we get $v_\star$ and $v_c$ from the GalPot package. For the Milky Way model \cite{2017MNRAS.465...76M} we consider in this paper, we have $v_c\lesssim 233 \, {\rm km/s}$. For $m_X$ in the range from $10^{-3} \, {\rm GeV}$ to $10^{3} \, {\rm GeV}$, we have $B\gtrsim 50$, then $\zeta$ can be approximated as $\zeta_\infty$, which only depends on $\eta$. For most of the NSs, we have $\eta\lesssim 2$, then $\zeta \gtrsim 0.44$.
For $m_X \ll 10^{-3} \, {\rm GeV} $ or $m_X \gg 10^3 \, {\rm GeV} $, then we have $B \to 0$ thus $\zeta \to \exp(-\eta^2)$, the suppression of the capture is even more significant.

To compare the amount of DM captured with and without the consideration of
Galactic motion, we introduce a parameter, the capture ratio $R_c$, which
estimates the ratio between the moving case and the corresponding static case.
The capture ratio is defined as, 
\begin{equation}
    R_c :=\frac{N_X^{({\rm m})}}{N_X^{({\rm s})}}
    =\frac{\int_0^{t_{\rm NS}} dt \
    C_X(t)   }{\int_0^{t_{\rm NS}} dt \ C_{X,0} } = \frac{\bar{A}}{A_{0}} \equiv \frac{ \frac{1}{t_{\rm NS}} \int_0^{t_{\rm NS}} dt \  \rho_{\rm DM}(t) \zeta(t) \bar{v}_{X,s}/ \bar{v}_X(t) }
    {\rho_{\rm DM,0} \bar{v}_{X,s}/\bar{v}_{X,0}} .
\end{equation}
Here, $N_X$ is the total number of DM particles captured by the NS, and the
superscripts ``$({\rm m})$'' and ``$({\rm s})$'' stand for the moving case and static case
respectively.    In the denominator, the NS is assumed to be static, which means we used the current
location of the NS to calculate $\rho_{\rm DM,0}$  and $\bar{v}_{X,0}$. Here, we introduce a typical velocity $\bar{v}_{X,s}=200 \, {\rm km/s}$ to make $\bar{A}$ and $A_0$ have the same dimension as $\rho_{\rm DM}$.

\begin{figure}
 \includegraphics[width=0.95\linewidth]{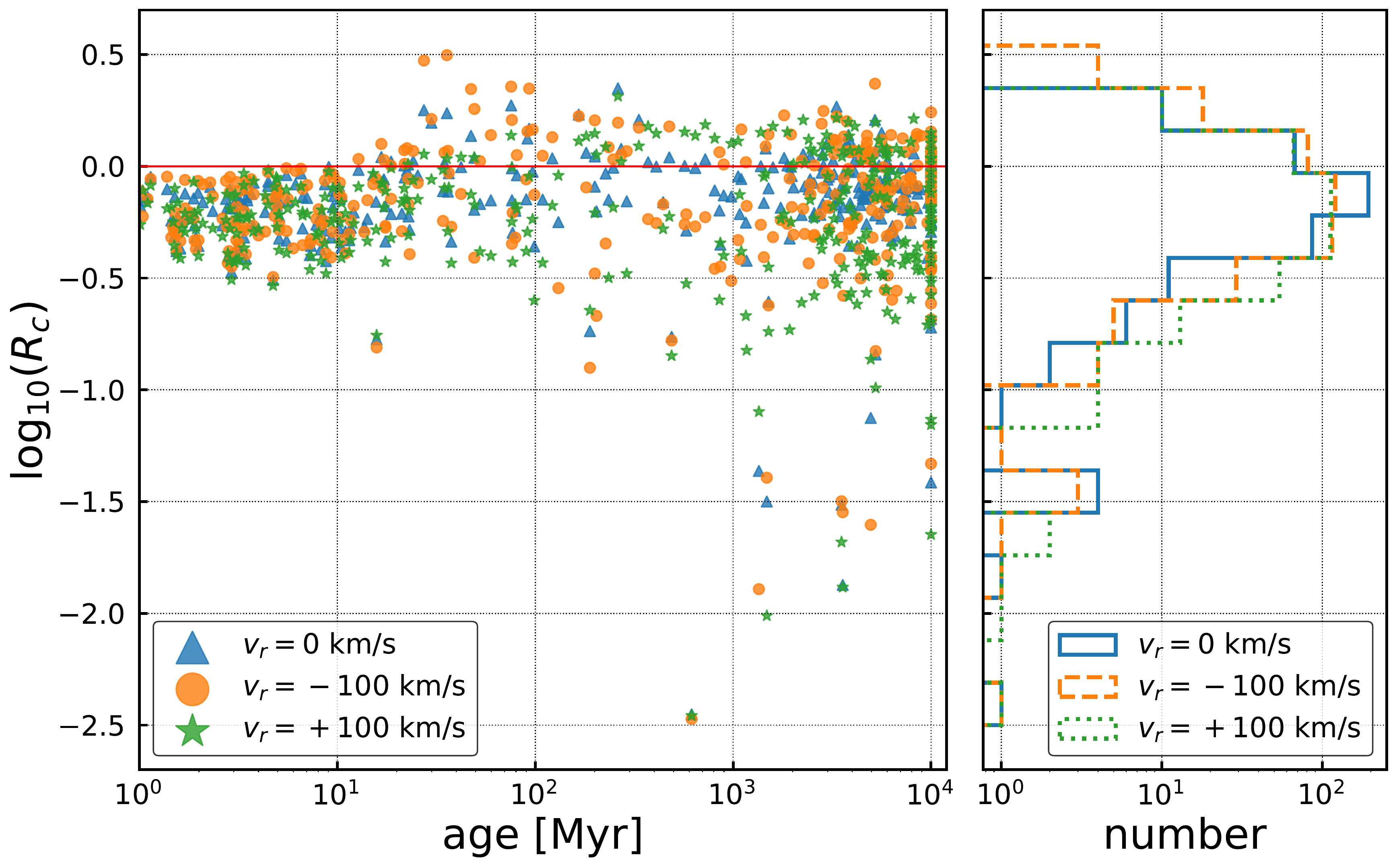}
 \caption{Distribution of the logarithm of the capture ratio $R_c$.  On the left
 panel, we show the distribution of $\log_{10}(R_c)$ and the age of 413 NSs.  We
 use different markers to denote three cases where the radial velocity $v_r=0,
 \, -100, \, +100 \, \rm{km/s}$.  On the right panel, we show the histograms of
 the numbers.  }
\label{Captureratio}
\end{figure}

Based on their Galactic motion, we calculated the capture ratio for 413 NSs, and
the results are shown in Fig.~\ref{Captureratio}. As we can see, the largest ratio is larger than 3 while the smallest ratio is about 0.003.  
Thus, after considering the Galactic motion, the capture amount of DM can change by about one to two orders of magnitude, compared to the static cases. 
For most of the NSs, the capture is suppressed due to the relative motion between the NSs and the DM halo. While for some of the NSs, $R_c>1$ since they have moved inwards to denser DM region where $\bar{v}_X$ is smaller and $\rho_{\rm DM}$ is much larger, so that the effect outstrips the suppression $\zeta$.

\section{Black Hole Formation}
\label{BHformation}

In this section, we consider three physical processes after the DM particles are
captured by the NS, namely, thermalization, collapse, and growth. These three
stages are important for the DM to eventually engulf the NS.

\subsection{Thermalization}

After being captured, the DM particles lose energy via repeated scattering with
nucleons and then attain thermal equilibrium with the star.  Ignoring nuclear
interactions and approximating the neutrons as a dense, non-interacting Fermi
gas, the timescale of thermalization can be estimated as \cite{Bertoni:2013bsa},
\begin{align}
    t_{{\rm th}} =& \frac{105\pi^2 \hbar^3 }{16 k_b^2T^2_{\rm NS} m_B \sigma_{nX}} \frac{m_X/m_B}{(1+m_X/m_B)^2}  
    = 75 \, {\rm yr} \, \bigg( \frac{ 10^{-45} \, {\rm cm}^2 }{ \sigma_{nX}} \bigg)\frac{m_X/m_B}{(1+m_X/m_B)^2} \, .
\end{align}
Here, $k_b$ is the Boltzmann constant.  To form a BH inside NS, we need that the
thermalization finishes within the lifetime of the NS, i.e.,
\begin{equation}
    t_{{\rm th}}<t_{{\rm NS}} \, .
    \label{CondThermal}
\end{equation}
We call Eq.~(\ref{CondThermal}) the ``Condition Thermalization.'' As the DM
thermalizes, it accumulates within a sphere of radius $r_{{\rm th}}$, which is
\begin{equation}
    r_{{\rm th}}=\sqrt{\frac{9k_b T_{{\rm NS}}}{8\pi G\rho_B m_X}} 
    = 8.8 \, {\rm m} \left( \frac{\rm GeV}{m_X} \right)^{1/2} ,
\end{equation}
where we have used the virial theorem
\begin{equation}
    \frac{G M_B(r_{{\rm th}})m_X}{r_{{\rm th}}} = \frac{3}{2}k_b T_{{\rm NS}} \, .
\end{equation}
We consider that the DM particles collect inside the NS, specifically, we
require the thermalization radius $r_{\rm th} \lesssim 0.2 R_{\rm NS}$
\cite{Bell:2013xk}, which gives a constraint that
\begin{equation}
    m_X \gtrsim \rm 1.6 \times 10^{-5} \, GeV.
\end{equation}

\subsection{Collapse}

To collapse into a BH, the accumulated DM has to form a self-gravitating system.
For bosonic DM, after it concentrates at the core of the NS, it can undergo a
phase transition to a BEC state, which has a higher density than a thermal gas-like
state.  Then the self-gravitation starts at a critical particle number
\begin{equation}
    N_X>N_{{\rm BEC}} \, .
\label{Selfgrav}
\end{equation}
For a Newtonian potential, McDermott et al.~\cite{McDermott:2011jp} provided the
derivation of $N_{\rm BEC}$, while for a compact object like NS, it is necessary
to consider the relativistic effects. After deriving an effective gravitational
potential energy, Jamison~\cite{Jamison:2013yya} gave the critical number for
BEC, that is
\begin{equation}
    N_{{\rm BEC}}=\zeta(3)\left(\frac{k_B T_{\rm NS}}{\hbar \sqrt{4\pi G (\rho_B+3P_B)/3}} \right)^3 = 
    5.6 \times 10^{38} \bigg( \frac{T_{\rm NS}}{10^6 \, \rm{K}} \bigg)^3 .
\label{Nbec}
\end{equation}
Here, we have adopted $P_B=0.3 \rho_B$ as in Ref.~\cite{Steiner:2012xt}. 
Since $N_{\rm BEC} \propto T^3_{\rm NS}$, the critical number for BEC reduces significantly for a cooler NS.

If we do not consider self-interaction, the critical number for Chandrasekhar
limit is $  N_{{\rm Cha}}=2 m^2_{\rm pl}/(\pi m^2_X) $, where $m_{\rm pl}$ is
the Planck mass \cite{Ruffini:1969qy}.  When we consider a bosonic field with a
self-coupling term $\lambda |\phi|^4$, the Chandrasekhar limit becomes
larger~\cite{Colpi:1986ye},
\begin{equation}
    N_{{\rm Cha}}=\frac{2m^2_{{\rm pl}}}{\pi m_X^2}\left(1+\frac{\lambda m_{{\rm pl}}^2}{32\pi m_X^2} \right)^{1/2} .
\label{ChaLimit}
\end{equation}
Such a repulsive self-interaction becomes stronger when $\lambda$ is larger. In
this paper, we adopt $\lambda=10^{-30}$ as in Refs.~\cite{Bramante:2013hn,Bramante:2014zca}.  The DM will collapse when 
\begin{equation}
    N_X-N_{\rm BEC}>N_{\rm Cha} .
\label{CondCollapse}
\end{equation}
We call Eq.~(\ref{CondCollapse}) as the ``Condition Collapse.'' Notice that
$N_{\rm Cha}$ monotonically decreases when $m_X$ increases, and we have
\begin{equation}
    N_{\rm Cha}\simeq \left\{ 
\begin{array}{cc}
   1.2\times 10^{41} \,  \left( \frac{{\rm GeV}}{m_X} \right)^3  ,  & \quad\quad  m_X \ll 10^3 \, \rm{GeV} , \\
  9.5\times 10^{37}  \,   \left( \frac{{\rm GeV}}{m_X} \right)^2 ,  & \quad\quad m_X \gg 10^3 \, \rm{GeV} .
\end{array}
\right.
\end{equation}
For $T_{\rm NS}=10^6 \, {\rm K}$, we have $N_{\rm Cha} =N_{\rm BEC}$ when $m_X =
5.9 \, {\rm GeV}$. Thus, the Condition Collapse becomes $N_X\gtrsim N_{\rm
Cha}$ for $m_X\ll 5.9 \, {\rm GeV}$, and  $N_X\gtrsim N_{\rm BEC}$ for $m_X\gg
5.9 \, {\rm GeV}$.

\subsection{Growth}

After its formation in the interior of the NS, the BH accretes the ordinary
matter inside the NS, which can be described by the Bondi accretion
\cite{1983bhwd.book.....S},
\begin{equation}
    \bigg(\frac{dM_{\rm{BH}}}{dt}\bigg)_{\rm{Bondi}}=
    \frac{4\pi \lambda_s \rho_B (G M_{\rm{BH}})^2}{v_s^3} \, .
\end{equation}
Here, $M_{\rm BH}= N_{\rm Cha}\, m_X$ is the mass of BH when it forms.  The
parameter $\lambda_s=0.25$ is the accretion eigenvalue for the transonic
solution, and $v_s$ is the sound speed in the NS, which we approximate to be
$0.17 \, c$ as in Ref.~\cite{Bell:2013xk}.  At the same time, the BH also
dissipates via Hawking radiation.  Thus, the BH will grow only when the
following condition satisfies,
\begin{equation}
    \frac{4\pi \lambda_s \rho_B (G M_{\rm{BH}})^2}{v_s^3}-\frac{\hbar c^4}{15360\pi (G M_{\rm{BH}})^2}+C_X m_X>0 \,,
\label{CondGrowth}
\end{equation}
where the second term corresponds to Hawking radiation. The third term corresponds to the accretion of the dark matter, since all newly captured particles go to the ground state directly after a BEC is formed \cite{McDermott:2011jp,Kouvaris:2010jy,Garani:2018kkd}. 
In other words, when the accretion of both the ordinary matter and the dark matter exceeds the Hawking radiation, the BH will grow. The continuous growth
of the BH will eventually destroy the NS. Here we call Eq.~(\ref{CondGrowth}) as
the ``Condition Growth.''

\section{Results and Discussions}
\label{discussions}

If the Conditions Thermalization, Collapse and Growth, i.e.,
Eqs.~(\ref{CondThermal}), (\ref{CondCollapse}) and (\ref{CondGrowth}), are
satisfied simultaneously, the DM will form a BH inside the NS and then engulf
the star.  Thus, we can use the current observational fact that these NSs exist
to exclude the parameter space where all these three conditions are met.

\begin{figure}
 \includegraphics[width=0.95\linewidth]{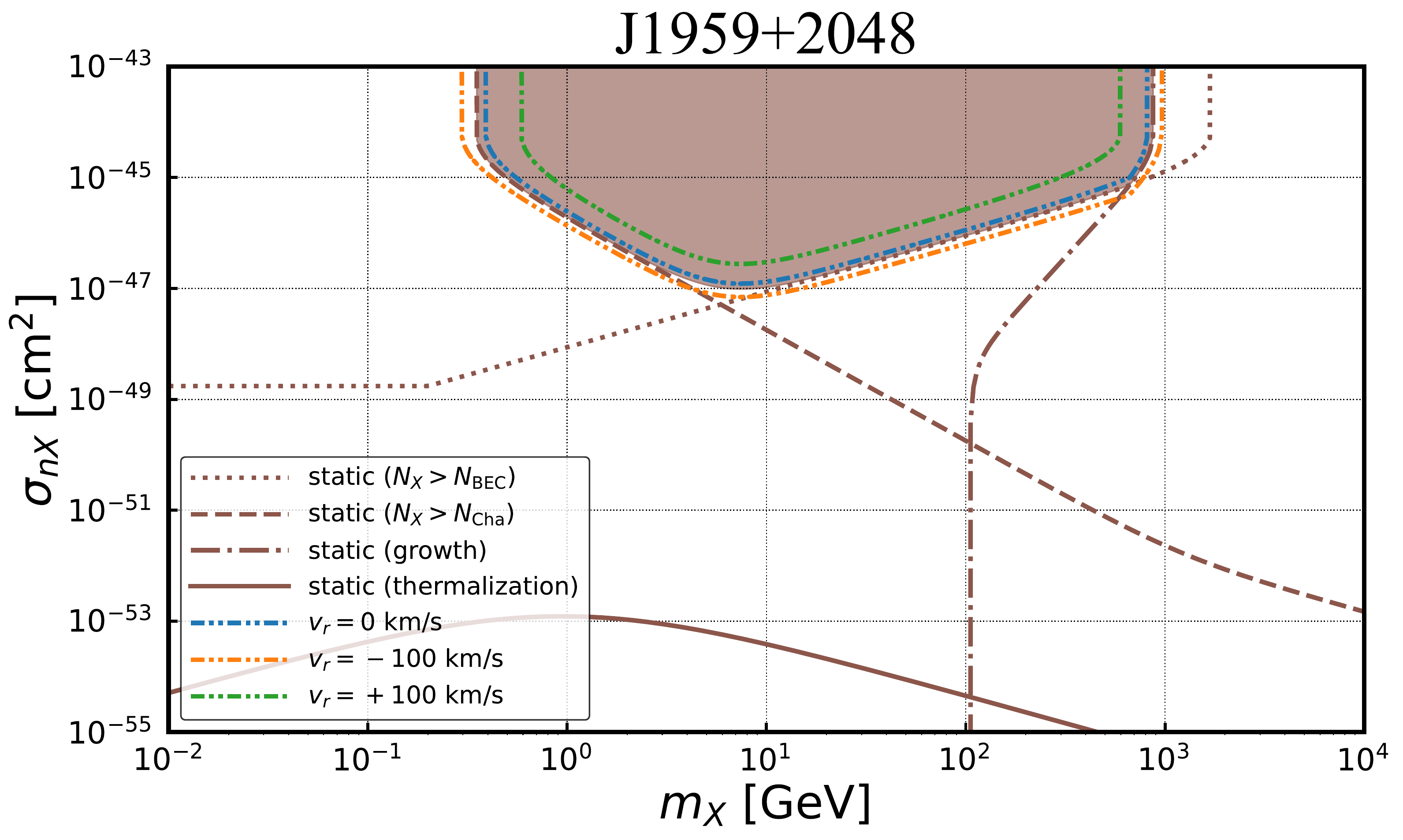}
 \caption{ Constraint on the DM-nucleon corss section $\sigma_{nX}$ from
 PSR~J1959+2048.  We use dotted, dashed, dot-dashed and solid lines in brown to
 denote Eqs.~(\ref{Selfgrav}), (\ref{ChaLimit}), (\ref{CondGrowth}) and
 (\ref{CondThermal}), respectively.  The shaded brown area represents the
 excluded parameter space due to the simultaneous satisfaction of the three
 conditions, that will deconstruct the NS.  The blue, orange, and green
 dot-dot-dashed lines correspond to the constraint in the cases where $v_r=0$,
 $-100 \, \rm{km/s}$, and $+100 \, \rm{km/s}$, respectively.
 }
\label{SCon}
\end{figure}

As we can see in Fig.~\ref{Galmotion}, PSR~J1959+2048 has quite different motion
for different choices of $v_r$.  For this NS, we have $R_c=0.80$, $1.39$, and
$0.35$,  for $v_r= 0$, $-100 \, \rm{km/s}$, and $+100 \, \rm{km/s}$,
respectively.  Thus, here we take it as an example  in Fig.~\ref{SCon}, to
illustrate the constraints in different scenarios.  For the static case, we use
brown lines with different styles to denote different conditions; the excluded
parameter space is shaded in light brown.  We also show the boundary of the
excluded parameter space in other three moving cases with dot-dot-dashed lines
in different colors.

Now, we analyze the constraint in details.  Considering
Eqs.~(\ref{capture}) and (\ref{NewCapture}), we have $1 \geq 1-(1-e^{-B^2})/B^2 \geq 0.8$ for the mass
range from $10^{-6} \, {\rm GeV}$ to $10^6 \, {\rm GeV}$, thus we approximate
this factor to be $1$ for the following estimation. If $\sigma_{nX} \ll \sigma_{\rm sat}$, we have
\begin{equation}
N_X \approx 2.1\times 10^{42}  
\bigg(\frac{\bar{A}}{{\rm GeV}/{\rm cm}^3}  \bigg)
\bigg( \frac{{t_{\rm NS}}}{\rm Gyr} \bigg)
\bigg( \frac{{\rm GeV}}{m_X} \bigg) \frac{\sigma_{nX}}{\sigma_{\rm sat}} \,.  
\end{equation}
For $ 5.9 \, {\rm GeV} > m_X >0.2 \, {\rm GeV}$, the Condition Collapse, namely
$N_X\gtrsim N_{\rm Cha}$, gives the bound on the cross section,
\begin{equation}
\sigma_{nX} \gtrsim 8.2 \times 10^{-47} \, {\rm cm}^2
\bigg(\frac{{\rm GeV}/{\rm cm}^3}{\bar{A}} \bigg)
\bigg( \frac{\rm Gyr}{{t_{\rm NS}}} \bigg)
\bigg( \frac{\rm GeV}{m_X} \bigg)^2 \,.
\label{sigmab1}
\end{equation}
For PSR~J1959$+$2048, it just corresponds to the left boundary of the constraint
curve (brown dashed line) in Fig.~\ref{SCon}, whose slope is $k\simeq -2$  in
the logarithm scale.  While for $m_X>5.9 \, {\rm GeV}$, the Condition Collapse,
namely  $N_X\gtrsim N_{\rm BEC}$, gives the bound,
\begin{equation}
\sigma_{nX} \gtrsim 4.0 \times 10^{-47} \, {\rm cm}^2
\bigg(\frac{{\rm GeV}/{\rm cm}^3}{\bar{A}} \bigg)
\bigg( \frac{\rm Gyr}{{t_{\rm NS}}} \bigg)
\bigg( \frac{m_X}{100 \, \rm GeV} \bigg) \,.
\label{sigmab2}
\end{equation}
It corresponds to the right boundary (brown dotted line)  in Fig.~\ref{SCon},
whose slope is $k \simeq 1$.  When $\sigma_{nX} \gg \sigma_{\rm sat}$, we have
$1- \exp(-\sigma_{nx}/\sigma_{\rm sat})  \to 1$.  For the lower mass range, we
have $N_{\rm Cha}>N_{\rm BEC}$, thus $N_X\gtrsim N_{\rm Cha}$ gives the lower
cutoff of the mass of the DM particle,
\begin{equation}
m_X \gtrsim 2.3 \times 10^{-1} \, {\rm GeV}
\bigg(\frac{{\rm GeV}/{\rm cm}^3}{\bar{A}} \bigg)^{1/2}
\bigg( \frac{\rm Gyr}{{t_{\rm NS}}}\bigg)^{1/2} \,,
\label{mxlow}
\end{equation}
while $N_X\gtrsim N_{\rm BEC}$ gives the higher cutoff,
\begin{equation}
    m_X \lesssim 3.8 \times 10^3 \, {\rm GeV}
\bigg(\frac{\bar{A}}{{\rm GeV}/{\rm cm}^3} \bigg)
\bigg( \frac{{t_{\rm NS}}}{\rm Gyr}\bigg) \,.
\label{mxhigh}
\end{equation}

For most NSs that are old enough, the Condition Thermalization is satisfied
trivially. Now we discuss the Condition Growth. Considering only the first two
terms in Eq.~(\ref{CondGrowth}), we get a conservative condition for the growth
of BH, 
\begin{equation}
M_{\rm BH} \geq \bigg( \frac{\hbar c^4 v_s^3 }{61440 \pi^2 G^4 \lambda_s \rho_B} \bigg)^{1/4} = 9.1 \times 10^{36} \, {\rm GeV} \, ,
\end{equation}
which corresponds to 
\begin{equation}
    m_X \leq 113 \, {\rm GeV} \,.
\end{equation}
It means that for $m_X \leq 113 \, {\rm GeV}$, the BH always grows after
formation, which is the lower cutoff for the dot-dashed brown curve in
Fig.~\ref{SCon}. While Eq.~(\ref{CondGrowth}) gives the higher cutoff, which is
$m_X \lesssim 931 \, {\rm GeV} \,\big[\bar{A}/(\rm{GeV/cm}^3)\big]^{1/2}$
for $\bar{A} \gg 7 \, \rm{GeV/cm}^3$ and $m_X \lesssim 1064 \, {\rm GeV}
\, \big[\bar{A}/(\rm{GeV/cm}^3)\big]^{1/4}$ for $\bar{A} \ll 7 \,
\rm{GeV/cm}^3$.  The boundary for $\sigma_{nX}$ is
\begin{equation}
    \sigma_{nX} \gtrsim 1.9 \times 10^{-48} {\rm cm}^2 \bigg(\frac{m_X}{200 \, {\rm GeV}} \bigg)^4 \bigg( \frac{{\rm GeV/cm}^3}{\bar{A} } \bigg) \,,
\label{sigmab3}
\end{equation}
when $m_X< 1000 \, {\rm GeV}$, thus the slope is $k\simeq 4$ (brown dot-dashed
line) in Fig.~\ref{SCon}.

\begin{figure}
 \includegraphics[width=\linewidth]{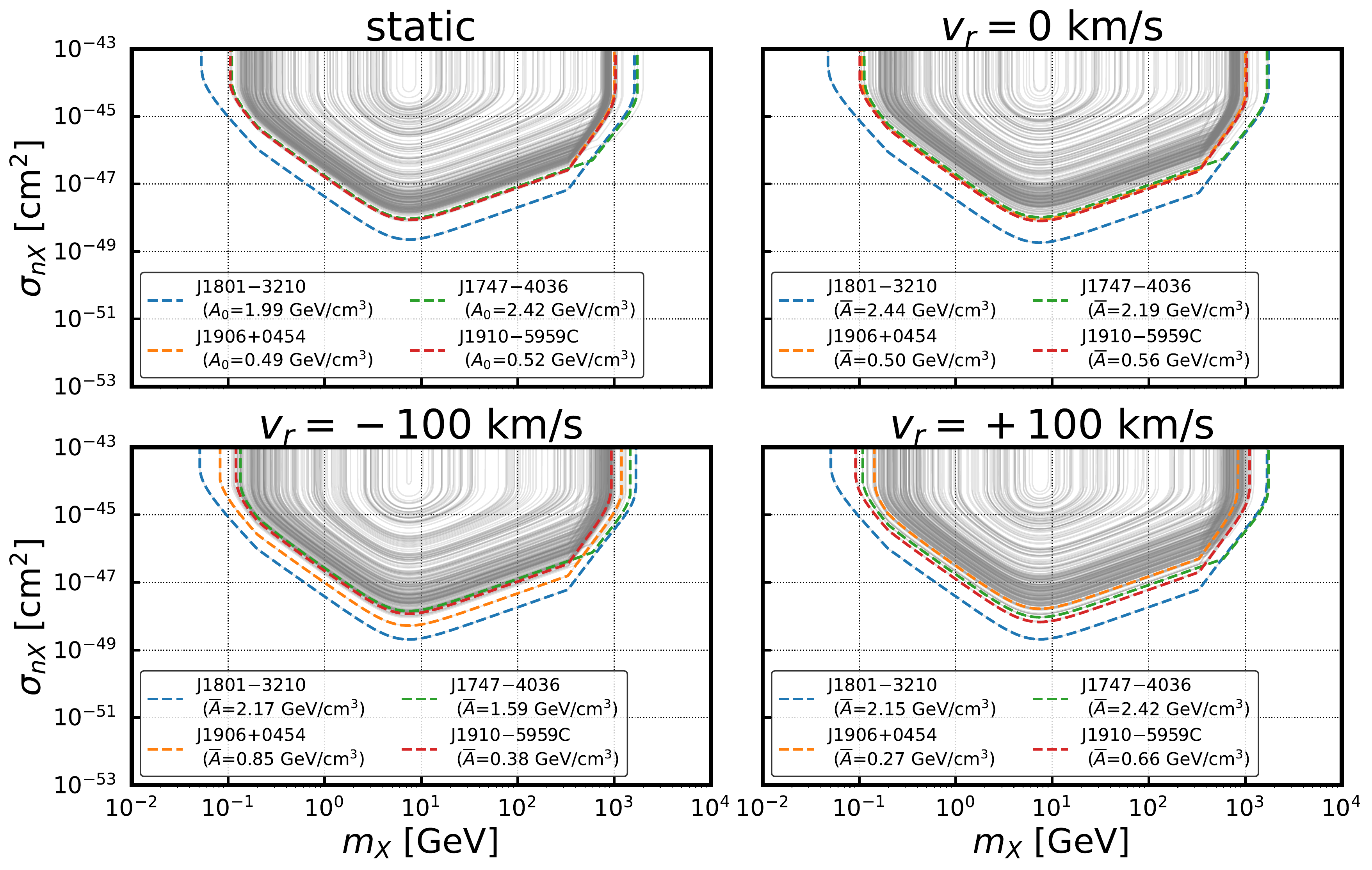}
 \caption{Constraints from 413 NSs. Each NS gives a thin grey line, and the
 parameter space above that line is excluded due to the survival of that NS.
 Overall, a few NSs contribute to the most outer bound, and these NSs are
 denoted  as thick dashed lines.}
\label{Constraint}
\end{figure}

From above analysis, we find that the boundary from Condition Collapse relies on
$\bar{A}$ and $t_{\rm NS}$, and the boundary from Condition Growth
only depends on $\bar{A}$.  Thus, the survival of those old NSs in
the center of Milky Way where the DM density is high, will give the most
stringent constraint.  Now, We use all the data from the 413 NSs to place a
combined constraint, and the results are shown in Fig.~\ref{Constraint}.  For
each NS, there is a thin grey line, representing the boundary of the excluded
parameter space. For the parameter space above the boundary, a BH will form and
destroy the star.  Thus, the outer envelope of these lines represents for the
combined constraint.  As we can see in Fig.~\ref{Constraint}, the most stringent
constraint comes from a few NSs, such as PSR~J1801$-$3210.  These few NSs are
highlighted with thick dashed lines.


To compare the moving cases with the static cases, we calculate the ratio
between the outer envelop of the combined constraint curve, namely,
\begin{equation}
    R_{\sigma}=\frac{\sigma_{nX}^{\rm (m)}}{\sigma_{nX}^{\rm (s)}} \,.
\end{equation}
As we can see in Eqs.~(\ref{sigmab1}), (\ref{sigmab2}) and (\ref{sigmab3}), the
boundary $\sigma_{nX} \propto 1/\bar{A}$, thus we have
\begin{equation}
    R_\sigma \simeq \frac{A_0}{\bar{A}}= \frac{1}{R_c} \, ,
\end{equation}
for the same $m_X$.  Since the boundary is primarily determined by
PSR~J1801$-$3210, we take it as an example.  We find that $R_\sigma= 0.81$,
$0.91$, and  $0.92$ for $v_r= 0$,  $-100 \, {\rm km/s}$, and $+100 \, {\rm
km/s}$, respectively.  The ratio $R_\sigma<1$  means that we can exclude the
smaller cross section for the moving cases compared to the static case. 
Notice that, the value of $\bar{A}$ varies case by case for the four highlighted NSs. It is majorly due to the different orbits in the past, and is influenced by the radial velocity, as shown in Fig.~\ref{Galmotion}. When we measure the radial velocity of pulsars in the future, we can construct more realistic orbits.

To conclude, for the first time we consider how the Galactic motion of NSs affects their accretion of DM. 
With the motion of the NSs, the DM density and velocity dispersion changes with time. The relative motion between the NSs and the DM halo suppresses the capture efficiency.
For some of the NSs, the capture amount can
change by as large as one to two orders of magnitude, compared to the static
cases.  We use the survival of NSs to constrain the mass of the DM particle and
the cross section. The envelope of constraints from 413 NSs gives an improved
bound.  In the future, we need more precise measurement of the radial velocity
to construct the precise trajectory of NSs. The age of NSs also needs to be
determined properly to place more realistic constraints.  
The effects of the age of NSs on the constraint on the cross section can be estimated by  Eqs.~(\ref{sigmab1}) and (\ref{sigmab2}).
In this work, we have only
considered bosonic DM particles, but a similar analysis can be applied to the
fermionic DM, which we leave to a future work.
Last but not least, we treat the NSs as density-uniform object and adopt the escape velocity at the surface of the star to simplify the calculation, as in Refs.~\cite{McDermott:2011jp, Bramante:2013nma, Bramante:2014zca}. While, in reality, the inner structure of NS is much more complicated, which also affects the capture of DM, as was indicated in Ref.~\cite{Bell:2020jou}. It will be another interesting topic in the future to discuss how different equation of states of NS affect the capture of NS. 

\begin{acknowledgments}
We acknowledge the referee for the useful comments which help us improve the manuscript.
We thank Zexin Hu, Xueli Miao, Zhongpeng Sun, Yichen Wang and Norbert Wex for
helpful discussion. This work was supported by the China Postdoctoral Science
Foundation (2021TQ0018), the National SKA Program of China (2020SKA0120300), 
the National Natural Science Foundation of China (11975027, 11991053), the Max 
Planck Partner Group Program funded by the Max Planck Society, and the 
High-Performance Computing Platform of Peking University.
\end{acknowledgments}


\begin{thebibliography}{100}
	
	\bibitem{Planck:2018vyg}
	{\scshape Planck} collaboration, \emph{{Planck 2018 results. VI. Cosmological
			parameters}},
	\href{https://doi.org/10.1051/0004-6361/201833910}{\emph{Astron. Astrophys.}
		{\bfseries 641} (2020) A6}
	[\href{https://arxiv.org/abs/1807.06209}{{\ttfamily 1807.06209}}].
	
	\bibitem{Bertone:2004pz}
	G.~Bertone, D.~Hooper and J.~Silk, \emph{{Particle dark matter: Evidence,
			candidates and constraints}},
	\href{https://doi.org/10.1016/j.physrep.2004.08.031}{\emph{Phys. Rept.}
		{\bfseries 405} (2005) 279}
	[\href{https://arxiv.org/abs/hep-ph/0404175}{{\ttfamily hep-ph/0404175}}].
	
	\bibitem{Clowe:2006eq}
	D.~Clowe, M.~Bradac, A.~H. Gonzalez, M.~Markevitch, S.~W. Randall, C.~Jones
	et~al., \emph{{A direct empirical proof of the existence of dark matter}},
	\href{https://doi.org/10.1086/508162}{\emph{Astrophys. J. Lett.} {\bfseries
			648} (2006) L109} [\href{https://arxiv.org/abs/astro-ph/0608407}{{\ttfamily
			astro-ph/0608407}}].
	
	\bibitem{Jungman:1995df}
	G.~Jungman, M.~Kamionkowski and K.~Griest, \emph{{Supersymmetric dark matter}},
	\href{https://doi.org/10.1016/0370-1573(95)00058-5}{\emph{Phys. Rept.}
		{\bfseries 267} (1996) 195}
	[\href{https://arxiv.org/abs/hep-ph/9506380}{{\ttfamily hep-ph/9506380}}].
	
	\bibitem{Feng:2010gw}
	J.~L. Feng, \emph{{Dark Matter Candidates from Particle Physics and Methods of
			Detection}},
	\href{https://doi.org/10.1146/annurev-astro-082708-101659}{\emph{Ann. Rev.
			Astron. Astrophys.} {\bfseries 48} (2010) 495}
	[\href{https://arxiv.org/abs/1003.0904}{{\ttfamily 1003.0904}}].
	
	\bibitem{Press:1985ug}
	W.~H. Press and D.~N. Spergel, \emph{{Capture by the sun of a galactic
			population of weakly interacting massive particles}},
	\href{https://doi.org/10.1086/163485}{\emph{Astrophys. J.} {\bfseries 296}
		(1985) 679}.
	
	\bibitem{Gould:1987ju}
	A.~Gould, \emph{{{WIMP} Distribution in and Evaporation From the Sun}},
	\href{https://doi.org/10.1086/165652}{\emph{Astrophys. J.} {\bfseries 321}
		(1987) 560}.
	
	\bibitem{Gould:1987ir}
	A.~Gould, \emph{{Resonant Enhancements in WIMP Capture by the Earth}},
	\href{https://doi.org/10.1086/165653}{\emph{Astrophys. J.} {\bfseries 321}
		(1987) 571}.
	
	\bibitem{Gould:1987ww}
	A.~Gould, \emph{{Direct and Indirect Capture of Wimps by the Earth}},
	\href{https://doi.org/10.1086/166347}{\emph{Astrophys. J.} {\bfseries 328}
		(1988) 919}.
	
	\bibitem{Kouvaris:2007ay}
	C.~Kouvaris, \emph{{WIMP Annihilation and Cooling of Neutron Stars}},
	\href{https://doi.org/10.1103/PhysRevD.77.023006}{\emph{Phys. Rev. D}
		{\bfseries 77} (2008) 023006}
	[\href{https://arxiv.org/abs/0708.2362}{{\ttfamily 0708.2362}}].
	
	\bibitem{Bertone:2007ae}
	G.~Bertone and M.~Fairbairn, \emph{{Compact Stars as Dark Matter Probes}},
	\href{https://doi.org/10.1103/PhysRevD.77.043515}{\emph{Phys. Rev. D}
		{\bfseries 77} (2008) 043515}
	[\href{https://arxiv.org/abs/0709.1485}{{\ttfamily 0709.1485}}].
	
	\bibitem{Kouvaris:2010vv}
	C.~Kouvaris and P.~Tinyakov, \emph{{Can Neutron stars constrain Dark Matter?}},
	\href{https://doi.org/10.1103/PhysRevD.82.063531}{\emph{Phys. Rev. D}
		{\bfseries 82} (2010) 063531}
	[\href{https://arxiv.org/abs/1004.0586}{{\ttfamily 1004.0586}}].
	
	\bibitem{McCullough:2010ai}
	M.~McCullough and M.~Fairbairn, \emph{{Capture of Inelastic Dark Matter in
			White Dwarves}},
	\href{https://doi.org/10.1103/PhysRevD.81.083520}{\emph{Phys. Rev. D}
		{\bfseries 81} (2010) 083520}
	[\href{https://arxiv.org/abs/1001.2737}{{\ttfamily 1001.2737}}].
	
	\bibitem{Bramante:2017xlb}
	J.~Bramante, A.~Delgado and A.~Martin, \emph{{Multiscatter stellar capture of
			dark matter}}, \href{https://doi.org/10.1103/PhysRevD.96.063002}{\emph{Phys.
			Rev. D} {\bfseries 96} (2017) 063002}
	[\href{https://arxiv.org/abs/1703.04043}{{\ttfamily 1703.04043}}].
	
	\bibitem{Baryakhtar:2017dbj}
	M.~Baryakhtar, J.~Bramante, S.~W. Li, T.~Linden and N.~Raj, \emph{{Dark Kinetic
			Heating of Neutron Stars and An Infrared Window On WIMPs, SIMPs, and Pure
			Higgsinos}},
	\href{https://doi.org/10.1103/PhysRevLett.119.131801}{\emph{Phys. Rev. Lett.}
		{\bfseries 119} (2017) 131801}
	[\href{https://arxiv.org/abs/1704.01577}{{\ttfamily 1704.01577}}].
	
	\bibitem{Raj:2017wrv}
	N.~Raj, P.~Tanedo and H.-B. Yu, \emph{{Neutron stars at the dark matter direct
			detection frontier}},
	\href{https://doi.org/10.1103/PhysRevD.97.043006}{\emph{Phys. Rev. D}
		{\bfseries 97} (2018) 043006}
	[\href{https://arxiv.org/abs/1707.09442}{{\ttfamily 1707.09442}}].
	
	\bibitem{Bell:2018pkk}
	N.~F. Bell, G.~Busoni and S.~Robles, \emph{{Heating up Neutron Stars with
			Inelastic Dark Matter}},
	\href{https://doi.org/10.1088/1475-7516/2018/09/018}{\emph{JCAP} {\bfseries
			09} (2018) 018} [\href{https://arxiv.org/abs/1807.02840}{{\ttfamily
			1807.02840}}].
	
	\bibitem{Camargo:2019wou}
	D.~A. Camargo, F.~S. Queiroz and R.~Sturani, \emph{{Detecting Dark Matter with
			Neutron Star Spectroscopy}},
	\href{https://doi.org/10.1088/1475-7516/2019/09/051}{\emph{JCAP} {\bfseries
			09} (2019) 051} [\href{https://arxiv.org/abs/1901.05474}{{\ttfamily
			1901.05474}}].
	
	\bibitem{Acevedo:2019agu}
	J.~F. Acevedo, J.~Bramante, R.~K. Leane and N.~Raj, \emph{{Warming Nuclear
			Pasta with Dark Matter: Kinetic and Annihilation Heating of Neutron Star
			Crusts}}, \href{https://doi.org/10.1088/1475-7516/2020/03/038}{\emph{JCAP}
		{\bfseries 03} (2020) 038}
	[\href{https://arxiv.org/abs/1911.06334}{{\ttfamily 1911.06334}}].
	
	\bibitem{Maity:2021fxw}
	T.~N. Maity and F.~S. Queiroz, \emph{{Detecting bosonic dark matter with
			neutron stars}},
	\href{https://doi.org/10.1103/PhysRevD.104.083019}{\emph{Phys. Rev. D}
		{\bfseries 104} (2021) 083019}
	[\href{https://arxiv.org/abs/2104.02700}{{\ttfamily 2104.02700}}].
	
	\bibitem{Nguyen:2022zwb}
	T.~T.~Q. Nguyen and T.~M.~P. Tait, \emph{{Bounds on Long-lived Dark Matter
			Mediators from Neutron Stars}},
	\href{https://arxiv.org/abs/2212.12547}{{\ttfamily 2212.12547}}.
	
	\bibitem{Chatterjee:2022dhp}
	S.~Chatterjee, R.~Garani, R.~K. Jain, B.~Kanodia, M.~S.~N. Kumar and S.~K.
	Vempati, \emph{{Faint light of old neutron stars from dark matter capture and
			detectability at the James Webb Space Telescope}},
	\href{https://arxiv.org/abs/2205.05048}{{\ttfamily 2205.05048}}.
	
	\bibitem{Sandin:2008db}
	F.~Sandin and P.~Ciarcelluti, \emph{{Effects of mirror dark matter on neutron
			stars}},
	\href{https://doi.org/10.1016/j.astropartphys.2009.09.005}{\emph{Astropart.
			Phys.} {\bfseries 32} (2009) 278}
	[\href{https://arxiv.org/abs/0809.2942}{{\ttfamily 0809.2942}}].
	
	\bibitem{Perez-Garcia:2010xlt}
	M.~A. Perez-Garcia, J.~Silk and J.~R. Stone, \emph{{Dark matter, neutron stars
			and strange quark matter}},
	\href{https://doi.org/10.1103/PhysRevLett.105.141101}{\emph{Phys. Rev. Lett.}
		{\bfseries 105} (2010) 141101}
	[\href{https://arxiv.org/abs/1007.1421}{{\ttfamily 1007.1421}}].
	
	\bibitem{Leung:2012vea}
	S.~C. Leung, M.~C. Chu and L.~M. Lin, \emph{{Equilibrium Structure and Radial
			Oscillations of Dark Matter Admixed Neutron Stars}},
	\href{https://doi.org/10.1103/PhysRevD.85.103528}{\emph{Phys. Rev. D}
		{\bfseries 85} (2012) 103528}
	[\href{https://arxiv.org/abs/1205.1909}{{\ttfamily 1205.1909}}].
	
	\bibitem{Li:2012qf}
	X.~Li, F.~Wang and K.~S. Cheng, \emph{{Gravitational effects of condensate dark
			matter on compact stellar objects}},
	\href{https://doi.org/10.1088/1475-7516/2012/10/031}{\emph{JCAP} {\bfseries
			10} (2012) 031} [\href{https://arxiv.org/abs/1210.1748}{{\ttfamily
			1210.1748}}].
	
	\bibitem{Perez-Garcia:2013dwa}
	M.~A. Perez-Garcia, F.~Daigne and J.~Silk, \emph{{Short GRBs and dark matter
			seeding in neutron stars}},
	\href{https://doi.org/10.1088/0004-637X/768/2/145}{\emph{Astrophys. J.}
		{\bfseries 768} (2013) 145}
	[\href{https://arxiv.org/abs/1303.2697}{{\ttfamily 1303.2697}}].
	
	\bibitem{Goldman:2013qla}
	I.~Goldman, R.~N. Mohapatra, S.~Nussinov, D.~Rosenbaum and V.~Teplitz,
	\emph{{Possible Implications of Asymmetric Fermionic Dark Matter for Neutron
			Stars}}, \href{https://doi.org/10.1016/j.physletb.2013.07.017}{\emph{Phys.
			Lett. B} {\bfseries 725} (2013) 200}
	[\href{https://arxiv.org/abs/1305.6908}{{\ttfamily 1305.6908}}].
	
	\bibitem{Xiang:2013xwa}
	Q.-F. Xiang, W.-Z. Jiang, D.-R. Zhang and R.-Y. Yang, \emph{{Effects of
			fermionic dark matter on properties of neutron stars}},
	\href{https://doi.org/10.1103/PhysRevC.89.025803}{\emph{Phys. Rev. C}
		{\bfseries 89} (2014) 025803}
	[\href{https://arxiv.org/abs/1305.7354}{{\ttfamily 1305.7354}}].
	
	\bibitem{Rezaei:2016zje}
	Z.~Rezaei, \emph{{Study of Dark-Matter Admixed Neutron Stars using the Equation
			of State from the Rotational Curves of Galaxies}},
	\href{https://doi.org/10.1088/1361-6528/aa5273}{\emph{Astrophys. J.}
		{\bfseries 835} (2017) 33}
	[\href{https://arxiv.org/abs/1612.02804}{{\ttfamily 1612.02804}}].
	
	\bibitem{Mukhopadhyay:2016dsg}
	S.~Mukhopadhyay, D.~Atta, K.~Imam, D.~N. Basu and C.~Samanta, \emph{{Compact
			bifluid hybrid stars: Hadronic Matter mixed with self-interacting fermionic
			Asymmetric Dark Matter}},
	\href{https://doi.org/10.1140/epjc/s10052-017-5006-3}{\emph{Eur. Phys. J. C}
		{\bfseries 77} (2017) 440}
	[\href{https://arxiv.org/abs/1612.07093}{{\ttfamily 1612.07093}}].
	
	\bibitem{Ellis:2018bkr}
	J.~Ellis, G.~H\"utsi, K.~Kannike, L.~Marzola, M.~Raidal and V.~Vaskonen,
	\emph{{Dark Matter Effects On Neutron Star Properties}},
	\href{https://doi.org/10.1103/PhysRevD.97.123007}{\emph{Phys. Rev. D}
		{\bfseries 97} (2018) 123007}
	[\href{https://arxiv.org/abs/1804.01418}{{\ttfamily 1804.01418}}].
	
	\bibitem{Das:2020vng}
	H.~C. Das, A.~Kumar, B.~Kumar, S.~Kumar~Biswal, T.~Nakatsukasa, A.~Li et~al.,
	\emph{{Effects of dark matter on the nuclear and neutron star matter}},
	\href{https://doi.org/10.1093/mnras/staa1435}{\emph{Mon. Not. Roy. Astron.
			Soc.} {\bfseries 495} (2020) 4893}
	[\href{https://arxiv.org/abs/2002.00594}{{\ttfamily 2002.00594}}].
	
	\bibitem{Das:2020ptd}
	H.~C. Das, A.~Kumar, B.~Kumar, S.~K. Biswal and S.~K. Patra, \emph{{Impacts of
			dark matter on the curvature of the neutron star}},
	\href{https://doi.org/10.1088/1475-7516/2021/01/007}{\emph{JCAP} {\bfseries
			01} (2021) 007} [\href{https://arxiv.org/abs/2007.05382}{{\ttfamily
			2007.05382}}].
	
	\bibitem{Kain:2021hpk}
	B.~Kain, \emph{{Dark matter admixed neutron stars}},
	\href{https://doi.org/10.1103/PhysRevD.103.043009}{\emph{Phys. Rev. D}
		{\bfseries 103} (2021) 043009}
	[\href{https://arxiv.org/abs/2102.08257}{{\ttfamily 2102.08257}}].
	
	\bibitem{Gleason:2022eeg}
	T.~Gleason, B.~Brown and B.~Kain, \emph{{Dynamical evolution of dark matter
			admixed neutron stars}},
	\href{https://doi.org/10.1103/PhysRevD.105.023010}{\emph{Phys. Rev. D}
		{\bfseries 105} (2022) 023010}
	[\href{https://arxiv.org/abs/2201.02274}{{\ttfamily 2201.02274}}].
	
	\bibitem{Shakeri:2022dwg}
	S.~Shakeri and D.~R. Karkevandi, \emph{{Bosonic Dark Matter in Light of the
			NICER Precise Mass-Radius Measurements}},
	\href{https://arxiv.org/abs/2210.17308}{{\ttfamily 2210.17308}}.
	
	\bibitem{Ellis:2017jgp}
	J.~Ellis, A.~Hektor, G.~H\"utsi, K.~Kannike, L.~Marzola, M.~Raidal et~al.,
	\emph{{Search for Dark Matter Effects on Gravitational Signals from Neutron
			Star Mergers}},
	\href{https://doi.org/10.1016/j.physletb.2018.04.048}{\emph{Phys. Lett. B}
		{\bfseries 781} (2018) 607}
	[\href{https://arxiv.org/abs/1710.05540}{{\ttfamily 1710.05540}}].
	
	\bibitem{Nelson:2018xtr}
	A.~Nelson, S.~Reddy and D.~Zhou, \emph{{Dark halos around neutron stars and
			gravitational waves}},
	\href{https://doi.org/10.1088/1475-7516/2019/07/012}{\emph{JCAP} {\bfseries
			07} (2019) 012} [\href{https://arxiv.org/abs/1803.03266}{{\ttfamily
			1803.03266}}].
	
	\bibitem{Kopp:2018jom}
	J.~Kopp, R.~Laha, T.~Opferkuch and W.~Shepherd, \emph{{Cuckoo\textquoteright{}s
			eggs in neutron stars: can LIGO hear chirps from the dark sector?}},
	\href{https://doi.org/10.1007/JHEP11(2018)096}{\emph{JHEP} {\bfseries 11}
		(2018) 096} [\href{https://arxiv.org/abs/1807.02527}{{\ttfamily
			1807.02527}}].
	
	\bibitem{Das:2018frc}
	A.~Das, T.~Malik and A.~C. Nayak, \emph{{Confronting nuclear equation of state
			in the presence of dark matter using GW170817 observation in relativistic
			mean field theory approach}},
	\href{https://doi.org/10.1103/PhysRevD.99.043016}{\emph{Phys. Rev. D}
		{\bfseries 99} (2019) 043016}
	[\href{https://arxiv.org/abs/1807.10013}{{\ttfamily 1807.10013}}].
	
	\bibitem{Alexander:2018qzg}
	S.~Alexander, E.~McDonough, R.~Sims and N.~Yunes, \emph{{Hidden-Sector
			Modifications to Gravitational Waves From Binary Inspirals}},
	\href{https://doi.org/10.1088/1361-6382/aaeb5c}{\emph{Class. Quant. Grav.}
		{\bfseries 35} (2018) 235012}
	[\href{https://arxiv.org/abs/1808.05286}{{\ttfamily 1808.05286}}].
	
	\bibitem{Quddus:2019ghy}
	A.~Quddus, G.~Panotopoulos, B.~Kumar, S.~Ahmad and S.~K. Patra, \emph{{GW170817
			constraints on the properties of a neutron star in the presence of WIMP dark
			matter}}, \href{https://doi.org/10.1088/1361-6471/ab9d36}{\emph{J. Phys. G}
		{\bfseries 47} (2020) 095202}
	[\href{https://arxiv.org/abs/1902.00929}{{\ttfamily 1902.00929}}].
	
	\bibitem{Horowitz:2019aim}
	C.~J. Horowitz and S.~Reddy, \emph{{Gravitational Waves from Compact Dark
			Objects in Neutron Stars}},
	\href{https://doi.org/10.1103/PhysRevLett.122.071102}{\emph{Phys. Rev. Lett.}
		{\bfseries 122} (2019) 071102}
	[\href{https://arxiv.org/abs/1902.04597}{{\ttfamily 1902.04597}}].
	
	\bibitem{Das:2021wku}
	H.~C. Das, A.~Kumar and S.~K. Patra, \emph{{Effects of dark matter on the
			in-spiral properties of the binary neutron stars}},
	\href{https://doi.org/10.1093/mnras/stab2387}{\emph{Mon. Not. Roy. Astron.
			Soc.} {\bfseries 507} (2021) 4053}
	[\href{https://arxiv.org/abs/2104.01815}{{\ttfamily 2104.01815}}].
	
	\bibitem{Das:2021yny}
	H.~C. Das, A.~Kumar and S.~K. Patra, \emph{{Dark matter admixed neutron star as
			a possible compact component in the GW190814 merger event}},
	\href{https://doi.org/10.1103/PhysRevD.104.063028}{\emph{Phys. Rev. D}
		{\bfseries 104} (2021) 063028}
	[\href{https://arxiv.org/abs/2109.01853}{{\ttfamily 2109.01853}}].
	
	\bibitem{Bezares:2019jcb}
	M.~Bezares, D.~Vigan\`o and C.~Palenzuela, \emph{{Gravitational wave signatures
			of dark matter cores in binary neutron star mergers by using numerical
			simulations}}, \href{https://doi.org/10.1103/PhysRevD.100.044049}{\emph{Phys.
			Rev. D} {\bfseries 100} (2019) 044049}
	[\href{https://arxiv.org/abs/1905.08551}{{\ttfamily 1905.08551}}].
	
	\bibitem{Das:2020ecp}
	A.~Das, T.~Malik and A.~C. Nayak, \emph{{Dark matter admixed neutron star
			properties in light of gravitational wave observations: A two fluid
			approach}}, \href{https://doi.org/10.1103/PhysRevD.105.123034}{\emph{Phys.
			Rev. D} {\bfseries 105} (2022) 123034}
	[\href{https://arxiv.org/abs/2011.01318}{{\ttfamily 2011.01318}}].
	
	\bibitem{Dengler:2021qcq}
	Y.~Dengler, J.~Schaffner-Bielich and L.~Tolos, \emph{{Second Love number of
			dark compact planets and neutron stars with dark matter}},
	\href{https://doi.org/10.1103/PhysRevD.105.043013}{\emph{Phys. Rev. D}
		{\bfseries 105} (2022) 043013}
	[\href{https://arxiv.org/abs/2111.06197}{{\ttfamily 2111.06197}}].
	
	\bibitem{Collier:2022cpr}
	M.~Collier, D.~Croon and R.~K. Leane, \emph{{Tidal Love numbers of novel and
			admixed celestial objects}},
	\href{https://doi.org/10.1103/PhysRevD.106.123027}{\emph{Phys. Rev. D}
		{\bfseries 106} (2022) 123027}
	[\href{https://arxiv.org/abs/2205.15337}{{\ttfamily 2205.15337}}].
	
	\bibitem{Karkevandi:2021ygv}
	D.~R. Karkevandi, S.~Shakeri, V.~Sagun and O.~Ivanytskyi, \emph{{Bosonic dark
			matter in neutron stars and its effect on gravitational wave signal}},
	\href{https://doi.org/10.1103/PhysRevD.105.023001}{\emph{Phys. Rev. D}
		{\bfseries 105} (2022) 023001}
	[\href{https://arxiv.org/abs/2109.03801}{{\ttfamily 2109.03801}}].
	
	\bibitem{Goldman:1989nd}
	I.~Goldman and S.~Nussinov, \emph{{Weakly Interacting Massive Particles and
			Neutron Stars}}, \href{https://doi.org/10.1103/PhysRevD.40.3221}{\emph{Phys.
			Rev. D} {\bfseries 40} (1989) 3221}.
	
	\bibitem{Gould:1989gw}
	A.~Gould, B.~T. Draine, R.~W. Romani and S.~Nussinov, \emph{{Neutron Stars:
			Graveyard of Charged Dark Matter}},
	\href{https://doi.org/10.1016/0370-2693(90)91745-W}{\emph{Phys. Lett. B}
		{\bfseries 238} (1990) 337}.
	
	\bibitem{Kouvaris:2018wnh}
	C.~Kouvaris, P.~Tinyakov and M.~H.~G. Tytgat, \emph{{NonPrimordial Solar Mass
			Black Holes}},
	\href{https://doi.org/10.1103/PhysRevLett.121.221102}{\emph{Phys. Rev. Lett.}
		{\bfseries 121} (2018) 221102}
	[\href{https://arxiv.org/abs/1804.06740}{{\ttfamily 1804.06740}}].
	
	\bibitem{Dasgupta:2020mqg}
	B.~Dasgupta, R.~Laha and A.~Ray, \emph{{Low Mass Black Holes from Dark Core
			Collapse}}, \href{https://doi.org/10.1103/PhysRevLett.126.141105}{\emph{Phys.
			Rev. Lett.} {\bfseries 126} (2021) 141105}
	[\href{https://arxiv.org/abs/2009.01825}{{\ttfamily 2009.01825}}].
	
	\bibitem{Garani:2021gvc}
	R.~Garani, D.~Levkov and P.~Tinyakov, \emph{{Solar mass black holes from
			neutron stars and bosonic dark matter}},
	\href{https://doi.org/10.1103/PhysRevD.105.063019}{\emph{Phys. Rev. D}
		{\bfseries 105} (2022) 063019}
	[\href{https://arxiv.org/abs/2112.09716}{{\ttfamily 2112.09716}}].
	
	\bibitem{Bhattacharya:2023stq}
	S.~Bhattacharya, B.~Dasgupta, R.~Laha and A.~Ray, \emph{{Can LIGO Detect
			Asymmetric Dark Matter?}},
	\href{https://arxiv.org/abs/2302.07898}{{\ttfamily 2302.07898}}.
	
	\bibitem{Guver:2012ba}
	T.~G\"uver, A.~E. Erkoca, M.~Hall~Reno and I.~Sarcevic, \emph{{On the capture
			of dark matter by neutron stars}},
	\href{https://doi.org/10.1088/1475-7516/2014/05/013}{\emph{JCAP} {\bfseries
			05} (2014) 013} [\href{https://arxiv.org/abs/1201.2400}{{\ttfamily
			1201.2400}}].
	
	\bibitem{Lopes:2020dau}
	J.~Lopes, T.~Lacroix and I.~Lopes, \emph{{Towards a more rigorous treatment of
			uncertainties on the velocity distribution of dark matter particles for
			capture in stars}},
	\href{https://doi.org/10.1088/1475-7516/2021/01/073}{\emph{JCAP} {\bfseries
			01} (2021) 073} [\href{https://arxiv.org/abs/2007.15927}{{\ttfamily
			2007.15927}}].
	
	\bibitem{Bell:2020obw}
	N.~F. Bell, G.~Busoni, T.~F. Motta, S.~Robles, A.~W. Thomas and M.~Virgato,
	\emph{{Nucleon Structure and Strong Interactions in Dark Matter Capture in
			Neutron Stars}},
	\href{https://doi.org/10.1103/PhysRevLett.127.111803}{\emph{Phys. Rev. Lett.}
		{\bfseries 127} (2021) 111803}
	[\href{https://arxiv.org/abs/2012.08918}{{\ttfamily 2012.08918}}].
	
	\bibitem{Bell:2020jou}
	N.~F. Bell, G.~Busoni, S.~Robles and M.~Virgato, \emph{{Improved Treatment of
			Dark Matter Capture in Neutron Stars}},
	\href{https://doi.org/10.1088/1475-7516/2020/09/028}{\emph{JCAP} {\bfseries
			09} (2020) 028} [\href{https://arxiv.org/abs/2004.14888}{{\ttfamily
			2004.14888}}].
	
	\bibitem{Bell:2020lmm}
	N.~F. Bell, G.~Busoni, S.~Robles and M.~Virgato, \emph{{Improved Treatment of
			Dark Matter Capture in Neutron Stars II: Leptonic Targets}},
	\href{https://doi.org/10.1088/1475-7516/2021/03/086}{\emph{JCAP} {\bfseries
			03} (2021) 086} [\href{https://arxiv.org/abs/2010.13257}{{\ttfamily
			2010.13257}}].
	
	\bibitem{Anzuini:2021lnv}
	F.~Anzuini, N.~F. Bell, G.~Busoni, T.~F. Motta, S.~Robles, A.~W. Thomas et~al.,
	\emph{{Improved treatment of dark matter capture in neutron stars III:
			nucleon and exotic targets}},
	\href{https://doi.org/10.1088/1475-7516/2021/11/056}{\emph{JCAP} {\bfseries
			11} (2021) 056} [\href{https://arxiv.org/abs/2108.02525}{{\ttfamily
			2108.02525}}].
	
	\bibitem{Bose:2022ola}
	D.~Bose and S.~Sarkar, \emph{{Impact of galactic distributions in celestial
			capture of dark matter}},  \href{https://arxiv.org/abs/2211.16982}{{\ttfamily
			2211.16982}}.
	
	\bibitem{Dasgupta:2019juq}
	B.~Dasgupta, A.~Gupta and A.~Ray, \emph{{Dark matter capture in celestial
			objects: Improved treatment of multiple scattering and updated constraints
			from white dwarfs}},
	\href{https://doi.org/10.1088/1475-7516/2019/08/018}{\emph{JCAP} {\bfseries
			08} (2019) 018} [\href{https://arxiv.org/abs/1906.04204}{{\ttfamily
			1906.04204}}].
	
	\bibitem{Dasgupta:2020dik}
	B.~Dasgupta, A.~Gupta and A.~Ray, \emph{{Dark matter capture in celestial
			objects: light mediators, self-interactions, and complementarity with direct
			detection}}, \href{https://doi.org/10.1088/1475-7516/2020/10/023}{\emph{JCAP}
		{\bfseries 10} (2020) 023}
	[\href{https://arxiv.org/abs/2006.10773}{{\ttfamily 2006.10773}}].
	
	\bibitem{Bertoni:2013bsa}
	B.~Bertoni, A.~E. Nelson and S.~Reddy, \emph{{Dark Matter Thermalization in
			Neutron Stars}},
	\href{https://doi.org/10.1103/PhysRevD.88.123505}{\emph{Phys. Rev. D}
		{\bfseries 88} (2013) 123505}
	[\href{https://arxiv.org/abs/1309.1721}{{\ttfamily 1309.1721}}].
	
	\bibitem{Garani:2020wge}
	R.~Garani, A.~Gupta and N.~Raj, \emph{{Observing the thermalization of dark
			matter in neutron stars}},
	\href{https://doi.org/10.1103/PhysRevD.103.043019}{\emph{Phys. Rev. D}
		{\bfseries 103} (2021) 043019}
	[\href{https://arxiv.org/abs/2009.10728}{{\ttfamily 2009.10728}}].
	
	\bibitem{Petraki:2013wwa}
	K.~Petraki and R.~R. Volkas, \emph{{Review of asymmetric dark matter}},
	\href{https://doi.org/10.1142/S0217751X13300287}{\emph{Int. J. Mod. Phys. A}
		{\bfseries 28} (2013) 1330028}
	[\href{https://arxiv.org/abs/1305.4939}{{\ttfamily 1305.4939}}].
	
	\bibitem{Zurek:2013wia}
	K.~M. Zurek, \emph{{Asymmetric Dark Matter: Theories, Signatures, and
			Constraints}},
	\href{https://doi.org/10.1016/j.physrep.2013.12.001}{\emph{Phys. Rept.}
		{\bfseries 537} (2014) 91} [\href{https://arxiv.org/abs/1308.0338}{{\ttfamily
			1308.0338}}].
	
	\bibitem{deLavallaz:2010wp}
	A.~de~Lavallaz and M.~Fairbairn, \emph{{Neutron Stars as Dark Matter Probes}},
	\href{https://doi.org/10.1103/PhysRevD.81.123521}{\emph{Phys. Rev. D}
		{\bfseries 81} (2010) 123521}
	[\href{https://arxiv.org/abs/1004.0629}{{\ttfamily 1004.0629}}].
	
	\bibitem{McDermott:2011jp}
	S.~D. McDermott, H.-B. Yu and K.~M. Zurek, \emph{{Constraints on Scalar
			Asymmetric Dark Matter from Black Hole Formation in Neutron Stars}},
	\href{https://doi.org/10.1103/PhysRevD.85.023519}{\emph{Phys. Rev. D}
		{\bfseries 85} (2012) 023519}
	[\href{https://arxiv.org/abs/1103.5472}{{\ttfamily 1103.5472}}].
	
	\bibitem{Kouvaris:2010jy}
	C.~Kouvaris and P.~Tinyakov, \emph{{Constraining Asymmetric Dark Matter through
			observations of compact stars}},
	\href{https://doi.org/10.1103/PhysRevD.83.083512}{\emph{Phys. Rev. D}
		{\bfseries 83} (2011) 083512}
	[\href{https://arxiv.org/abs/1012.2039}{{\ttfamily 1012.2039}}].
	
	\bibitem{Kouvaris:2012dz}
	C.~Kouvaris and P.~Tinyakov, \emph{{(Not)-constraining heavy asymmetric bosonic
			dark matter}}, \href{https://doi.org/10.1103/PhysRevD.87.123537}{\emph{Phys.
			Rev. D} {\bfseries 87} (2013) 123537}
	[\href{https://arxiv.org/abs/1212.4075}{{\ttfamily 1212.4075}}].
	
	\bibitem{Fan:2012qy}
	Y.-z. Fan, R.-z. Yang and J.~Chang, \emph{{Constraining Asymmetric Bosonic
			Non-interacting Dark Matter with Neutron Stars}},
	\href{https://arxiv.org/abs/1204.2564}{{\ttfamily 1204.2564}}.
	
	\bibitem{Kouvaris:2013kra}
	C.~Kouvaris and P.~Tinyakov, \emph{{Growth of Black Holes in the interior of
			Rotating Neutron Stars}},
	\href{https://doi.org/10.1103/PhysRevD.90.043512}{\emph{Phys. Rev. D}
		{\bfseries 90} (2014) 043512}
	[\href{https://arxiv.org/abs/1312.3764}{{\ttfamily 1312.3764}}].
	
	\bibitem{East:2019dxt}
	W.~E. East and L.~Lehner, \emph{{Fate of a neutron star with an endoparasitic
			black hole and implications for dark matter}},
	\href{https://doi.org/10.1103/PhysRevD.100.124026}{\emph{Phys. Rev. D}
		{\bfseries 100} (2019) 124026}
	[\href{https://arxiv.org/abs/1909.07968}{{\ttfamily 1909.07968}}].
	
	\bibitem{Kouvaris:2011fi}
	C.~Kouvaris and P.~Tinyakov, \emph{{Excluding Light Asymmetric Bosonic Dark
			Matter}}, \href{https://doi.org/10.1103/PhysRevLett.107.091301}{\emph{Phys.
			Rev. Lett.} {\bfseries 107} (2011) 091301}
	[\href{https://arxiv.org/abs/1104.0382}{{\ttfamily 1104.0382}}].
	
	\bibitem{Jamison:2013yya}
	A.~O. Jamison, \emph{{Effects of gravitational confinement on bosonic
			asymmetric dark matter in stars}},
	\href{https://doi.org/10.1103/PhysRevD.88.035004}{\emph{Phys. Rev. D}
		{\bfseries 88} (2013) 035004}
	[\href{https://arxiv.org/abs/1304.3773}{{\ttfamily 1304.3773}}].
	
	\bibitem{Kouvaris:2011gb}
	C.~Kouvaris, \emph{{Limits on Self-Interacting Dark Matter}},
	\href{https://doi.org/10.1103/PhysRevLett.108.191301}{\emph{Phys. Rev. Lett.}
		{\bfseries 108} (2012) 191301}
	[\href{https://arxiv.org/abs/1111.4364}{{\ttfamily 1111.4364}}].
	
	\bibitem{Gresham:2018rqo}
	M.~I. Gresham and K.~M. Zurek, \emph{{Asymmetric Dark Stars and Neutron Star
			Stability}}, \href{https://doi.org/10.1103/PhysRevD.99.083008}{\emph{Phys.
			Rev. D} {\bfseries 99} (2019) 083008}
	[\href{https://arxiv.org/abs/1809.08254}{{\ttfamily 1809.08254}}].
	
	\bibitem{Garani:2022quc}
	R.~Garani, M.~H.~G. Tytgat and J.~Vandecasteele, \emph{{Condensed dark matter
			with a Yukawa interaction}},
	\href{https://doi.org/10.1103/PhysRevD.106.116003}{\emph{Phys. Rev. D}
		{\bfseries 106} (2022) 116003}
	[\href{https://arxiv.org/abs/2207.06928}{{\ttfamily 2207.06928}}].
	
	\bibitem{Bramante:2013hn}
	J.~Bramante, K.~Fukushima and J.~Kumar, \emph{{Constraints on bosonic dark
			matter from observation of old neutron stars}},
	\href{https://doi.org/10.1103/PhysRevD.87.055012}{\emph{Phys. Rev. D}
		{\bfseries 87} (2013) 055012}
	[\href{https://arxiv.org/abs/1301.0036}{{\ttfamily 1301.0036}}].
	
	\bibitem{Bramante:2013nma}
	J.~Bramante, K.~Fukushima, J.~Kumar and E.~Stopnitzky, \emph{{Bounds on
			self-interacting fermion dark matter from observations of old neutron
			stars}}, \href{https://doi.org/10.1103/PhysRevD.89.015010}{\emph{Phys. Rev.
			D} {\bfseries 89} (2014) 015010}
	[\href{https://arxiv.org/abs/1310.3509}{{\ttfamily 1310.3509}}].
	
	\bibitem{Bell:2013xk}
	N.~F. Bell, A.~Melatos and K.~Petraki, \emph{{Realistic neutron star
			constraints on bosonic asymmetric dark matter}},
	\href{https://doi.org/10.1103/PhysRevD.87.123507}{\emph{Phys. Rev. D}
		{\bfseries 87} (2013) 123507}
	[\href{https://arxiv.org/abs/1301.6811}{{\ttfamily 1301.6811}}].
	
	\bibitem{Garani:2018kkd}
	R.~Garani, Y.~Genolini and T.~Hambye, \emph{{New Analysis of Neutron Star
			Constraints on Asymmetric Dark Matter}},
	\href{https://doi.org/10.1088/1475-7516/2019/05/035}{\emph{JCAP} {\bfseries
			05} (2019) 035} [\href{https://arxiv.org/abs/1812.08773}{{\ttfamily
			1812.08773}}].
	
	\bibitem{Ray:2023auh}
	A.~Ray, \emph{{Celestial Objects as Strongly-Interacting Asymmetric Dark Matter
			Detectors}},  \href{https://arxiv.org/abs/2301.03625}{{\ttfamily
			2301.03625}}.
	
	\bibitem{Bramante:2014zca}
	J.~Bramante and T.~Linden, \emph{{Detecting Dark Matter with Imploding Pulsars
			in the Galactic Center}},
	\href{https://doi.org/10.1103/PhysRevLett.113.191301}{\emph{Phys. Rev. Lett.}
		{\bfseries 113} (2014) 191301}
	[\href{https://arxiv.org/abs/1405.1031}{{\ttfamily 1405.1031}}].
	
	\bibitem{Liu:2020hkx}
	K.~Liu et~al., \emph{{A revisit of PSR J1909\ensuremath{-}3744 with 15-yr
			high-precision timing}},
	\href{https://doi.org/10.1093/mnras/staa2993}{\emph{Mon. Not. Roy. Astron.
			Soc.} {\bfseries 499} (2020) 2276}
	[\href{https://arxiv.org/abs/2009.12544}{{\ttfamily 2009.12544}}].
	
	\bibitem{2016ascl.soft11006M}
	P.~J. {McMillan}, ``{GalPot: Galaxy potential code}.'' Astrophysics Source Code
	Library, record ascl:1611.006, Nov., 2016.
	
	\bibitem{Dehnen:1996fa}
	W.~Dehnen and J.~Binney, \emph{{Mass models of the Milky Way}},
	\href{https://doi.org/10.1046/j.1365-8711.1998.01282.x}{\emph{Mon. Not. Roy.
			Astron. Soc.} {\bfseries 294} (1998) 429}
	[\href{https://arxiv.org/abs/astro-ph/9612059}{{\ttfamily
			astro-ph/9612059}}].
	
	\bibitem{McMillan:2011wd}
	P.~J. McMillan, \emph{{Mass models of the Milky Way}},
	\href{https://doi.org/10.1111/j.1365-2966.2011.18564.x}{\emph{Mon. Not. Roy.
			Astron. Soc.} {\bfseries 414} (2011) 2446}
	[\href{https://arxiv.org/abs/1102.4340}{{\ttfamily 1102.4340}}].
	
	\bibitem{2017MNRAS.465...76M}
	P.~J. {McMillan}, \emph{{The mass distribution and gravitational potential of
			the Milky Way}}, \href{https://doi.org/10.1093/mnras/stw2759}{\emph{Mon. Not.
			Roy. Astron. Soc.} {\bfseries 465} (2017) 76}
	[\href{https://arxiv.org/abs/1608.00971}{{\ttfamily 1608.00971}}].
	
	\bibitem{Manchester:2004bp}
	R.~N. Manchester, G.~B. Hobbs, A.~Teoh and M.~Hobbs, \emph{{The Australia
			Telescope National Facility pulsar catalogue}},
	\href{https://doi.org/10.1086/428488}{\emph{Astron. J.} {\bfseries 129}
		(2005) 1993} [\href{https://arxiv.org/abs/astro-ph/0412641}{{\ttfamily
			astro-ph/0412641}}].
	
	\bibitem{Tauris:2012jp}
	T.~M. Tauris, N.~Langer and M.~Kramer, \emph{{Formation of millisecond pulsars
			with CO white dwarf companions - II. Accretion, spin-up, true ages and
			comparison to MSPs with He white dwarf companions}},
	\href{https://doi.org/10.1111/j.1365-2966.2012.21446.x}{\emph{Mon. Not. Roy.
			Astron. Soc.} {\bfseries 426} (2012) 1601}
	[\href{https://arxiv.org/abs/1206.1862}{{\ttfamily 1206.1862}}].
	
	\bibitem{Navarro:1996gj}
	J.~F. Navarro, C.~S. Frenk and S.~D. White, \emph{{A Universal density profile
			from hierarchical clustering}},
	\href{https://doi.org/10.1086/304888}{\emph{Astrophys. J.} {\bfseries 490}
		(1997) 493} [\href{https://arxiv.org/abs/astro-ph/9611107}{{\ttfamily
			astro-ph/9611107}}].
	
	\bibitem{Green:2010gw}
	A.~M. Green, \emph{{Dependence of direct detection signals on the WIMP velocity
			distribution}},
	\href{https://doi.org/10.1088/1475-7516/2010/10/034}{\emph{JCAP} {\bfseries
			10} (2010) 034} [\href{https://arxiv.org/abs/1009.0916}{{\ttfamily
			1009.0916}}].
	
	\bibitem{Catena:2011kv}
	R.~Catena and P.~Ullio, \emph{{The local dark matter phase-space density and
			impact on WIMP direct detection}},
	\href{https://doi.org/10.1088/1475-7516/2012/05/005}{\emph{JCAP} {\bfseries
			05} (2012) 005} [\href{https://arxiv.org/abs/1111.3556}{{\ttfamily
			1111.3556}}].
	
	\bibitem{Strigari:2012acq}
	L.~E. Strigari, \emph{{Galactic Searches for Dark Matter}},
	\href{https://doi.org/10.1016/j.physrep.2013.05.004}{\emph{Phys. Rept.}
		{\bfseries 531} (2013) 1} [\href{https://arxiv.org/abs/1211.7090}{{\ttfamily
			1211.7090}}].
	
	\bibitem{Steiner:2012xt}
	A.~W. Steiner, J.~M. Lattimer and E.~F. Brown, \emph{{The Neutron Star
			Mass-Radius Relation and the Equation of State of Dense Matter}},
	\href{https://doi.org/10.1088/2041-8205/765/1/L5}{\emph{Astrophys. J. Lett.}
		{\bfseries 765} (2013) L5} [\href{https://arxiv.org/abs/1205.6871}{{\ttfamily
			1205.6871}}].
	
	\bibitem{Ruffini:1969qy}
	R.~Ruffini and S.~Bonazzola, \emph{{Systems of selfgravitating particles in
			general relativity and the concept of an equation of state}},
	\href{https://doi.org/10.1103/PhysRev.187.1767}{\emph{Phys. Rev.} {\bfseries
			187} (1969) 1767}.
	
	\bibitem{Colpi:1986ye}
	M.~Colpi, S.~L. Shapiro and I.~Wasserman, \emph{{Boson Stars: Gravitational
			Equilibria of Selfinteracting Scalar Fields}},
	\href{https://doi.org/10.1103/PhysRevLett.57.2485}{\emph{Phys. Rev. Lett.}
		{\bfseries 57} (1986) 2485}.
	
	\bibitem{1983bhwd.book.....S}
	S.~L. {Shapiro} and S.~A. {Teukolsky}, \emph{{Black holes, white dwarfs, and
			neutron stars : the physics of compact objects}}. 1983.
	
\end{thebibliography}

\providecommand{\href}[2]{#2}\begingroup\raggedright\endgroup

\end{document}